\documentclass[conference,10pt]{IEEEtran}
\IEEEoverridecommandlockouts

\usepackage{cite}
\usepackage{amsmath,amssymb,amsfonts}
\usepackage{algorithmic}
\usepackage{graphicx}
\usepackage{url}
\usepackage{textcomp}
\usepackage{xcolor}
\def\BibTeX{{\rm B\kern-.05em{\sc i\kern-.025em b}\kern-.08em
    T\kern-.1667em\lower.7ex\hbox{E}\kern-.125emX}}
\begin{document}

\title{Offline Reinforcement Learning \\ for Mobility Robustness Optimization
}

\author{\IEEEauthorblockN{Pegah Alizadeh$^*$}
\IEEEauthorblockA{\textit{\hspace{+0cm}Ericsson Research\hspace{+0cm}} \\
Massy, France \\
}
\and
\IEEEauthorblockN{Anastasios Giovanidis$^*$}
\IEEEauthorblockA{\textit{\hspace{+0cm}Ericsson Research\hspace{+0cm}} \\
Massy, France 
}
\and
\IEEEauthorblockN{Pradeepa Ramachandra}
\IEEEauthorblockA{\textit{\hspace{+0cm}Ericsson Research\hspace{+0cm}} \\
Link{\"o}ping, Sweden \\
}
\and
\IEEEauthorblockN{Vasileios Koutsoukis}
\IEEEauthorblockA{\textit{\hspace{+0cm}Ericsson Research\hspace{+0cm}} \\
Massy, France 
}
\and
\IEEEauthorblockN{Osama Arouk}
\IEEEauthorblockA{\textit{\hspace{+0cm}Ericsson R{\&}D\hspace{+0cm}} \\
Massy, France 
}
}

\maketitle

\begin{abstract}
In this work\let\thefootnote\relax\footnotetext{$^*$ Equal contribution. e-mail: \{firstname.lastname\}@ericsson.com} we revisit the Mobility Robustness Optimisation (MRO) algorithm and study the possibility of learning the optimal Cell Individual Offset tuning using offline Reinforcement Learning. Such methods make use of collected offline datasets to learn the optimal policy, without further exploration. We adapt and apply a sequence-based method called Decision Transformers as well as a value-based method called Conservative Q-Learning to learn the optimal policy for the same target reward as the vanilla rule-based MRO. The same input features related to failures, ping-pongs, and other handover issues are used. Evaluation for realistic New Radio networks with 3500 MHz carrier frequency on a traffic mix including diverse user service types and a specific tunable cell-pair shows that offline-RL methods outperform rule-based MRO, offering up to $7\%$ improvement. Furthermore, offline-RL can be trained for diverse objective functions using the same available dataset, thus offering operational flexibility compared to rule-based methods.
\end{abstract}


\section{Introduction}
Self-Organizing Network (SON) functionalities have become key aspects of modern cellular networks for automation. They make use of collected data to allow the network to self-configure, self-optimize, and self-heal. Mobility Robustness Optimization (MRO) is the related SON feature whose
aim is to optimize the configuration of relevant mobility parameters and allow users to experience seamless connectivity. Users are physically moving from one cell to another with various speeds and trajectories. The network governs this geographical shift of service, on the same frequency, through the handover (HO) procedure. There are a number of tunable HO parameters (Time-To-Trigger, offsets, hysteresis, Cell Individual Offset, etc) that determine when a HO from the source to the target cell is triggered. The main focus of MRO is on the configuration of the Cell Individual Offset between adjacent cell pairs, i.e. $CIO_{i,j}$. Rule-based versions of MRO have already been deployed on LTE networks and have proved essential for optimizing handover performance and improving overall network efficiency~\cite{NgKwKi18}. 

In commercial networks, MRO operates centrally in a loop 
as follows: data from network configuration and performance management Key Performance Indicators (KPIs) are collected within a time window. These are often related to HO successes, failures and other events. 
Using a function of these measurements, the rule-based MRO algorithm decides, in comparison with some tunable thresholds to be explained later in the paper, whether to keep or modify the current CIO value to follow the evolution of traffic pattern. Such reconfiguration reduces Radio Link Failures (RLFs) and unnecessary handovers. 

In 5G New Radio (NR) and next-generation networks, artificial intelligence-based solutions can assist the network to 
automatically and proactively change configurations to quickly adapt to evolving environments. 
Early on, Reinforcement Learning (RL) algorithms were used. The motivation has been to learn the optimal HO parameter configuration based on online experience data, collected in the current cell measurements. This enables MRO to more effectively adapt to network traffic dynamics and improve generalization. In approaches using Q-learning the observation state can be very broad to include the cell load \cite{MaBeSp23} and the user velocity \cite{MwMi14}. The MRO was further extended to output different configurations per service type and Quality-of-Experience (QoE), such as the E-MRO~\cite{MaMwRa21} and more recently the slice-aware MRO~\cite{LiHuWe22}. These approaches not only enrich the state-space with service-quality information but also introduce more general and flexible reward functions.  

\subsection{Challenges and Contributions}

Training RL agents through interaction with real environments can be both expensive and risky. In MRO, for instance, an agent selects a cell individual offset (CIO) value for a specific cell pair, which is held constant over tens of minutes to gather sufficient handover performance data. Given that RL typically requires thousands of iterations to converge, online training would take several months. Additionally, exploration during training may involve suboptimal configurations, risking degraded network performance and user experience.

To mitigate these risks, RL is often pre-trained on realistic simulators or digital twins before being fine-tuned on real-world data. However, simulators used in industry for LTE or NR networks are extremely detailed, modeling full protocol stacks (from PHY and MAC to TCP and application layers) alongside realistic UE mobility and service dynamics. While this high fidelity enables safe testing, it also results in slow simulation speeds. For instance, in our setup, simulating a single 300-second interval of network time with mobility can take several hours and depends on the simulated traffic load.
This computational burden makes online RL training infeasible. Using digital twins as a possible workaround also raises challenges similar to a simulator, since their precision and hence their execution speed is in relation to the detail they are built to represent the physical network. They can also use real data for calibration. 


Motivated by these challenges, we address the CIO tuning problem in this work using \textit{offline reinforcement learning}. These algorithms learn entirely from previously collected data, gathered in parallel across multiple sources (simulators, measurements), without requiring step-by-step online exploration. This approach significantly reduces the cost and time associated with training while ensuring safety in action selection, because no potentially harmful decisions are made in a live environment. Offline datasets can be collected once, either from detailed simulators or from real networks where traditional MRO mechanisms are already deployed.
%
The same datasets can be used as benchmarks to train various RL algorithms that can fulfill different and more flexible objectives, without the need to collect new data each time. A similar approach in the context of link adaptation has been explored in  \cite{offlineLinkAdaptation}.

This paper is organized as follows: Section \ref{sec:MRO} presents the A3 event \cite{TS38331}, governed by a set of mobility control parameters, that triggers intra-frequency handover. In the same section, the taxonomy of issues due to suboptimal parameter configuration, as well as the existing rule-based MRO algorithm that we use as a reference, are presented. Next, Section \ref{sec:O-RL} introduces the RL formulation and presents the two offline RL algorithms that we apply, i.e., Decision Transformers (DT) \cite{DT21} and Conservative Q-Learning (CQL) \cite{CQL20}. The choice of these two specific methods will be further clarified. We refer to the proposed RL-based MRO algorithm in this work as \textit{offline-MRO} (either DT or CQL). 
The evaluation and comparison of offline-MRO with the rule-based MRO is presented in Section \ref{sec:evaluation} on a realistic 5G NR simulator with various UE service types, loads and velocities. Conclusions are drawn in Section \ref{sec:conclusions}.

\section{System model}
\label{sec:MRO}

\subsection{Intra-frequency handover}

\begin{figure}[t!]
    \centering
    \includegraphics[width=.5\textwidth]{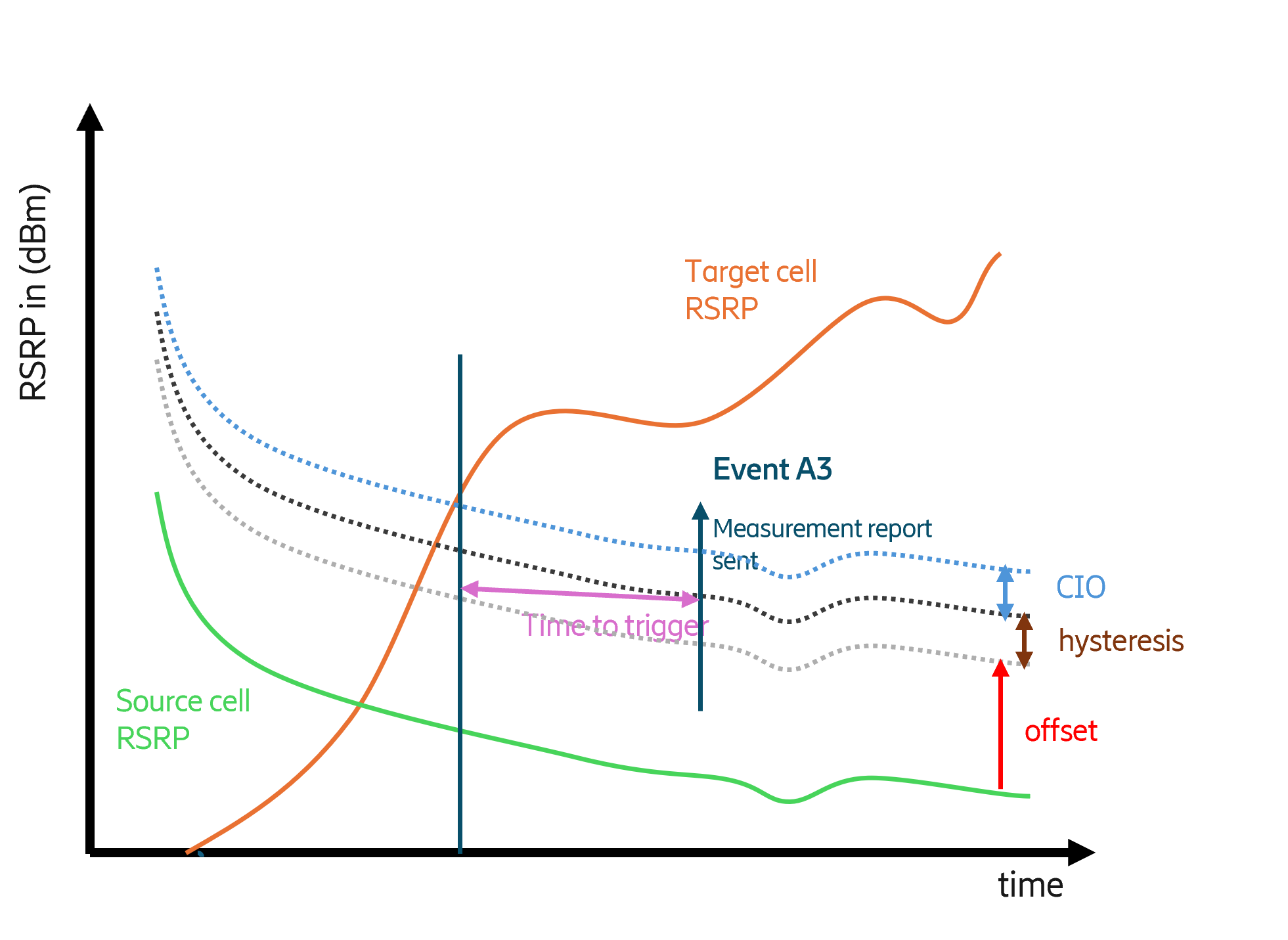}
    \caption{A3 event for intra-frequency handover.}
    \label{fig:A3}
\end{figure}

For a mobile user in the network, the process of handover from one cell to another on the same frequency band is triggered by the A3 event (see \cite{TS38331}, par. 5.5.4.4). The process is illustrated in Fig.~\ref{fig:A3}. On the UE side, the Reference Signal Received Power (RSRP) values from the source cell and neighboring cell (called target) are measured. Once the target signal becomes stronger than the source signal plus an offset,  Time-To-Trigger (TTT) supervision timer starts. If this difference persists for the period determined by the TTT, an A3 event is considered fulfilled. The UE then sends a measurement report to the serving cell, requesting a handover to the target cell. 
The A3 event entry condition based on the signal strength is 
\begin{eqnarray}
\label{rsrp_a3}
M_{t}-M_{s} > Off_{A3} + Hys_{A3} + CIO_{s,t},
\end{eqnarray}
where $M_t, M_s$ are the signal strength measurements at the UE from target and source cells, respectively, $Off_{A3}$ is the global offset, $Hys_{A3}$ is the hysteresis and $CIO_{s,t}$ is the Cell Individual Offset value, when moving from source cell to target. Note that in some expressions the convention is to use $- CIO_{s,t}$ instead of $+ CIO_{s,t}$, in  equation~\ref{rsrp_a3}). This does not affect generality, as the same results can be obtained by simply multiplying the CIO by $-1$.
The A3 offset and hysteresis values are common across all cells operating on the same frequency.
%
The pair-adjacent CIO is a local HO parameter that can be tuned individually (i.e. for each cell pair), in order to offer flexibility based on local coverage conditions. 
We focus in this work on how to optimally tune a single cell-pair. 

\subsection{Handover Performance Events}
\label{ho-events}

Suboptimal configuration of the cell-pair CIO from source cell $i$ to target cell $j$ can lead to undesirable effects. Here we present a list of issues due to suboptimal configuration of the cell-pair $(i,j)=(A,B)$ only. These will be used later for better tuning of the specific $CIO_{AB}$. 

\begin{itemize}
\item \textit{RLF Too Early HO (FTE)}: (\textit{early issue}) The A3 event is triggered too early because the pair CIO value is configured low. As a result, the signal level from $B$ is not sufficient to establish connection and an RLF occurs. The UE later re-establishes connection to source cell $A$.

\item \textit{RLF Too Late HO (FTL)}: (\textit{late issue}) The UE stays too long in the source cell $A$, to the point that the signal becomes too weak, because the pair CIO is configured high. This results in RLF at any point before the handover command 
is sent. The UE later  reconnects to cell $B$.

\item \textit{Ping-Pong (PP)}: (\textit{early issue}) The handover from $A$ to $B$ is successful, but is immediately followed by another successful HO from $B$ to $A$. The time interval between the two successful HOs is small, usually $1$ sec. 

\item \textit{Short stay Early-event (SE)}: (\textit{early issue}) The HO from $A$ to $B$ is successful, followed immediately by a successful HO to another cell $C$. 

\item \textit{Short stay Late-event (SL)}: (\textit{late issue}) The HO from $A$ to some other cell $C$ is successful, followed  immediately by a successful HO from $C$ to cell $B$. 

\item \textit{Success-then-Fail (StF)}: (\textit{early issue}) HO from $A$ to $B$ is successful, followed by a HO effort towards cell $C$ and RLF immediately after, within a short interval ($1$ sec). 

\item \textit{HO to Wrong Cell (WC)}: (\textit{early issue}) HO from cell $A$ to $B$ fails and the UE re-establishes connection to some other cell $C$. This means $B$ was the wrong cell to hand over in the first place. 

\item \textit{HO to Wrong Cell Re-establishment (RC)}: (\textit{late issue}) The UE aims to re-establish connection to cell $B$ after a failed HO from $A$ to some cell $C$. This means the cell $C$ was the wrong cell, and the HO should have been realized from $A$ to $B$ in the first place.
\end{itemize}


\subsection{Mobility Robustness Optimization}

MRO is a SON functionality that automatically tunes the CIO of a cell pair when the current configuration results in a high occurrence of HO issues. 
Its implementation in LTE networks has a simple rule-based decision-making logic~\cite{NgKwKi18,JoAhJo20}. During a predefined time-window, the system gathers handover statistics, by counting the successful, failed and redundant HOs. A function of early and late issues is proposed to quantify each part:
\begin{eqnarray}
\label{early-sum}
E_{sum} & = & W_fN_{FTE} + W_{p}N_{PP} +\nonumber\\ 
 & + &  [W_wN_{WC} + W_{ss}N_{StF} + 
W_{ss}N_{SE}] \\
\label{late-sum}
L_{sum} & = & W_fN_{FTL} + 
[W_wN_{RC} + W_{ss}N_{SL}].
\end{eqnarray}
In the above $N_{(event)}$ is the number of measured events for each type of handover $event$. Furthermore, the number of each event is multiplied by a weight to quantify its importance. Specifically, $W_f$ is the weight for failures, $W_p$ for ping-pongs, $W_w$ for the wrong cell, and $W_{ss}$ for short-stay related events. In practice, failures are more serious than the wrong cell, than ping-pongs, than short-stays $W_f>W_w>W_p>W_{ss}$. These weights are empirically selected to guide the algorithm, for example $W_f=1$, $W_w = 0.5$, $W_{p}=W_{ss}=0.1$. 



The reason we consider early and late events separately is that the two quantities $(E_{sum},L_{sum})$ (representing the number of early and late handovers, respectively) have opposite behaviour when changing the CIO. Specifically, when increasing the CIO, $E_{sum}$ decreases monotonically, whereas $L_{sum}$ increases, and vice-versa (see also \cite{CaVaGe23}). This gives rise to a simple MRO that adjusts CIO by $+1$ or $-1$ based on the relative values of these quantities.
The variations considered as the baseline in this work keeps early and late issue counts within some margin of each other.
To tune the MRO operation point, it uses the combined ratio,
\begin{eqnarray}
\label{ratio_MRO}
\rho_{MRO} & = & \frac{E_{sum}-L_{sum}}{N_{ALL}}.
\end{eqnarray}
\textbf{Rule-based MRO.} 
The MRO algorithm relies on three thresholds $\tau_{events}, \tau_{early}, \tau_{late}$ similarly to those used in the algorithms in~\cite{NgKwKi18,JoAhJo20} and others in the literature. The procedure in the presented MRO variation is as follows: Once the total number of events is sufficient, a decision rule (C.R) is evaluated: $-\tau_{late}<\rho_{MRO}<\tau_{early}$. If this condition is satisfied, the CIO value remains unchanged. If $\rho_{MRO}>\tau_{early}$, this implies that the number of early issues exceeds 
$\tau_{early}N_{ALL}$ by a greater margin than the late issues,
so we update $CIO\leftarrow CIO+1$. Otherwise, if $\rho_{MRO}<-\tau_{late}$, then 
we update $CIO\leftarrow CIO-1$. A choice of margins can be $\tau_{early}=\tau_{late}=1\%$.

The method is limited to select as operational point the CIO that balances early and late handover events; its success strongly depends on the opposite monotone behavior of the two. This does not allow for generalizations to alternative objectives that mix the issues differently, or are related to other KPIs, such as those associated to QoS, throughput, interruption times, etc. 
Furthermore, it does not allow to include new features as input (such as cell load, UE speed or neighbor cell information) thus limiting the application range capabilities of rule-based MRO. It has been shown in the literature~\cite{MwMi14, MaMwRa21, LiHuWe22} that, RL-based approaches to MRO do not suffer from such drawbacks and are more flexible. As argued in the introduction we will learn the optimal CIO using offline-RL methods.

\section{Offline-RL for pairwise CIO tuning}
\label{sec:O-RL}

We consider the problem of tuning the CIO for a directional pair of cells as a sequential decision-making problem over a long horizon $t=0,1,\ldots$. 
The environment can be described as a Markov Decision Process (MDP) by the tuple $(\mathcal{S},\mathcal{A},\mathcal{T},\mathcal{R},\gamma)$.

\textit{States:} $\mathcal{S}$ is the space of all possible states $s$. 
During the $\Delta \tau$ decision interval, the system counts HO-related events, as the ones described in Section \ref{ho-events}. 
We propose to include the indexed $1-11$ features in the state, as shown in Table \ref{tab:state-features}.

\begin{table}[h!]
\caption{State features.}
\label{tab:state-features}
\centering
\begin{tabular}{| c | c | c |}
\hline
Index & Feature & details\\
\hline\hline
1 & $CIO_{AB}$ & current cio value\\
\hline
2 & $N_{SUC}$ & successful HO count\\
\hline
3,4,5 & $N_{FTE}$, $N_{FTL}$,  $N_F$ & early, late, total RLF counts\\
\hline
6 & $N_{PP}$ & ping-pong counts\\
\hline
7,8 & $N_{SE}$, $N_{SL}$ & short-stay early and late counts\\
\hline
9 & $N_{StF}$ & success-then-fail count\\
\hline
10,11 & $N_{WC}$, $N_{RC}$ & wrong cell, re-establish counts\\
\hline
\end{tabular}
\end{table}

\textit{Actions:} The action space $\mathcal{A}=\{-1,0,+1\}$ allows the current CIO of the cell pair, to either increase by $+1\ dB$ or stay the same, or decrease by $-1\ dB$. We allow incremental modification, following the example of rule-based MRO algorithms, in order to guarantee gradual system updates that do not destabilize UE performance. 
Furthermore, the CIO value of the cell pair is allowed to vary between $CIO\in\{-8,\ldots,+8\}\ dB$. Notice that the specifications \cite{TS38331} allow for a larger range \{-24,\ldots,+24\} but in practice this is unsafe for service. 

\textit{Transition probabilities:} The probability of the environment moving to a new state $s'$ given the current one $s$ and an action $a$ is $\mathcal{T}:\mathcal{S}\times \mathcal{A}\times \mathcal{S}\rightarrow [0,1]$. The new state will consist of the updated counts of HO-related events, due to the updated CIO value. The transition is stochastic due to randomness in traffic (UE mobility patterns) and channel. 

\textit{Reward:} The reward in our case is a function $r:\mathcal{S}\rightarrow \mathbb{R}_+$  of the next state $s'$. Our goal is to formulate a reward function for the RL algorithm that aligns with the objective of the rule-based MRO, enabling a fair comparison between the two approaches. Recall that the rule-based MRO employed the ratio-condition in equation (\ref{ratio_MRO}) to maintain $E_{sum}$ and $L_{sum}$ as balanced as possible. In fact this balancing also works as a practical heuristic approximating the optimal $CIO^*$ that minimizes the total HO issues, i.e. $E_{sum}(CIO)+L_{sum}(CIO)$ in (\ref{full-cost}).  Rule-based MRO needs the ratio-test to decide how to modify the CIO value for lack of other way to combine the available data, whereas the RL-approach can learn directly the optimal action that minimizes the sum of HO issues. 

As a first step in designing the RL reward, we define the cost per state as a weighted sum of the \textit{rates} of early and late handover issues. To do so, we sum up the expressions in equations (\ref{early-sum}) and (\ref{late-sum}), and then divide them 
by the number of total handover events $N_{ALL}=N_{SUC}+(N_{FTE}+N_{FTL})$, in order to obtain rates. This follows the same logic as the rule-based MRO algorithm in equation (\ref{ratio_MRO}), while deliberately excluding short-stay events and wrong-cell handovers, which are already accounted for within the success and failure counts.  
We then define the reward $r(s)$ as the exponential of the negative cost plus some constant $C>0$.
\begin{eqnarray}
    \label{full-cost}
    c(s) & = & 
    w_{early}\frac{E_{sum}(s)}{N_{ALL}(s)} + w_{late}\frac{L_{sum}(s)}{N_{ALL}(s)}\\
    \label{full-reward}
    r(s) & = & \exp(C-c(s)).
\end{eqnarray}
 In expression (\ref{full-reward}), the reward is maximized at the same CIO value that minimizes the cost $c(s)$, since $\exp(-x)$ is a monotonic and convex function. The constant $C$ calibrates the range of rewards. 
The weights $w_{early}, w_{late}$ are chosen to favor early or late issues, depending on the application (here $w_{early}=w_{late}=1$). 

In the RL approach, alternative objectives can be proposed that shift the operational point for the MRO. For example, the operator may want to avoid strong deviation from the global offset value. This can be achieved by introducing a soft constraint in the reward, i.e.
\begin{eqnarray}
\label{p-cio}
r_{cio}(s) & = & r(s) -\lambda_{cio}|CIO|^{\alpha},
\end{eqnarray}
with $\lambda_{cio}>0$, and exponent $\alpha\geq 1$. Alternatively, the operator may want to avoid unnecessary handovers and introduce a linear penalty of the total number of handovers times a price (e.g. due to service interruption) $r(s)-\lambda_{event}N_{ALL}(s)$. Many other linear or non-linear alternatives can be introduced in RL, possibly also including extra information. 


\textit{RL-Trajectories:} An RL-trajectory\footnote{Not to be confused with UE physical trajectories due to mobility. These types of trajectories are frequently used in the concept of offline RL.} from some initial state $s_0$ to a final state $s_T$ consists of a sequence of states, actions, and rewards $d= (s_0, a_0, r_0, s_1, a_1, r_1,\ldots, s_T, a_T, r_T )$ up to a finite horizon (number of steps) $T$. Specifically, the instantaneous reward $r_t = r(s_t)$ is defined in (\ref{full-reward}) (or variations of it like (\ref{p-cio})) and is calculated as a function of the system state $s_t$ at time-step $t$. 
The action at time-step $t$ is taken following some predefined policy, given the current state $s_t$ and maybe past history. 
The next state $s_{t+1}$ consists of all counts in Table \ref{tab:state-features} as measured within the interval $\Delta \tau$ after tuning $CIO_{AB}$ (action) at the beginning of the time-step $t$. 

\textit{Reward-to-Go (RtG):} 
It is defined as the cumulative undiscounted return $R=\sum_{t=1}^T r_t$, which quantifies how much is achieved in an RL-trajectory of length $T$. This notion of return was first used in the context of Decision Transformers~\cite{DT21}. 

We are looking for an optimal CIO adjustment
policy $\pi(a|s)$ to maximize the expected RtG 
$\max_{\pi} \mathbb{E}^{\pi}\left[\sum_{t=0}^{T} r_t\right]$.
We include in the reward each step of the update and not just the final converged state $s_T$, because in a real system, the MRO algorithm will work continuously, adjusting the CIO by $+1$ or $-1$ depending on the HO-issue measured statistics due to traffic mobility. Hence, every time step with its HO-issues is important. Even in the special case where traffic conditions remain the same and only the converged value of $CIO$ is important, the choice to maximize over the sum of rewards will eventually force the algorithm to converge to the optimal tuning of $CIO$ (i.e. the one with the fewest HO-issues) as early as possible. In the objective to be maximized, the expectation over the RtG is due to the stochasticity of the environment (UE mobility, channels), assuming that the policy is deterministic.

\subsection{Offline-RL for MRO}

As mentioned in the introduction, online RL has certain limitations, especially for SON methods like MRO which need several tens of minutes in order to collect KPIs and evaluate the outcome of each single action. 
Another issue may arise due to online exploration, which could probe actions that can be dangerous for the offered service. For these reasons, as mentioned, we study the alternative to train the method with offline RL using the available collected datasets.


In our study we choose to work with two algorithms that have shown excellence in the literature: DT (Decision Transformers) \cite{DT21}, and CQL (Conservative Q-Learning) \cite{CQL20} for offline RL. Our choice results from the conclusions drawn in \cite{DTvsCQL}, which compared these two algorithms and showed that they both are  competitive but can outperform one another in different types of environments. Specifically, DT proves to be more robust than CQL when more offline data is available. However, CQL excels in situations of high stochasticity.

\textit{Offline dataset:} 
In our offline RL setting we consider there exists a dataset $\mathcal{D}$ of trajectories $d\in\mathcal{D}$, each having fixed length $T>0$. The trajectories are collected either through simulations, or from online probing. 
These trajectories have been collected for specific instances of cell traffic, e.g. characterized by a cell load $\ell_A$ and average velocity $v_A$. 
The collected trajectories can follow specific sub-optimal policies $\pi(a|s)$ and here is an indicative list:
\begin{itemize}
\item RND: Random actions $\pi(a|s):$ $a \sim Uniform(\mathcal{A})$. 
\item UP: Increasing CIO $\pi(a|s): a=+1$
\item DOWN: Decreasing CIO $\pi(a|s): a=-1$
\item MRO: Variations of the MRO $\pi(a|s): a \sim MRO(s,\mathcal{A})$ \end{itemize}
The suboptimal policies RND, UP and DOWN 
serve a valuable role in exploring the 
diversity of possible state-action sequences. 
The MRO variations can be tested on real systems for tuning and can provide high-reward expert trajectories.

\subsection{Conservative Q-Learning (CQL)}
CQL is a value-based method. Its main idea is to avoid overestimation of Out-Of-Distribution (OOD) actions, by minimizing the Q-values of unseen or less frequently visited state-action pairs, while at the same time maximizing the values of those pairs that appear in the offline dataset~\cite{CQL20}. We apply here the CQL version for discrete actions, 
\begin{eqnarray}
& \mathcal{L}(\theta) = \mathcal{L}_{DoubleDQN}(\theta) + \hspace{+4cm}\nonumber \\
& \alpha  \mathbb{E}_{s_t \in D}[\; \log \sum_{a \in \mathcal{A}} \exp{Q_{\theta}(s_t, a)} - \mathbb{E}_{a \in D} [Q_{\theta}(s_t,a)] \; ] 
\end{eqnarray}
The first line in the loss is the standard Double DQN loss. 
The second line in the loss corresponds to the CQL penalty term. It penalizes the difference between the Q-values of all possible actions $a \in \mathcal{A}$ and those observed in the dataset $a \in D$. 
The higher the coefficient $\alpha$, the more this difference is penalized. We use the d3rlpy package and adapt its CQL implementation to our problem\footnote{\url{https://d3rlpy.readthedocs.io/en/v2.7.0/}}.

\subsection{Decision Transformers}

A great breakthrough in offline RL came with the use of transformers, which allows the task to be cast as a sequence modeling problem, where the next action is the output from a causally masked transformer. Specifically, \cite{DT21} introduces the Decision Transformer (DT), which produces future actions in an autoregressive way, by conditioning on target Rewards-to-Go (RtG).  
Its efficiency lies in the utilization of the attention mechanism of transformers that can quantify the relation of the next outcome with the past states, actions, rewards. 
DT has the ability to \textit{stitch} together parts from suboptimal trajectories in the dataset and produce an optimal one. In our case, DT takes as input per decision step the last $K\geq 1$ timesteps with three inputs each: (a) the reward-to-go $R$, (b) the state $s$, and (c) the action $a$. 
The value $K$ is called \textit{context-size} and is an important hyperparameter that determines the history taken into account from the transformer. 
We adapt the HuggingFace implementation to our problem
\footnote{\url{ https://github.com/huggingface/transformers}} by involving the cross-entropy loss during training to predict the discrete actions.

\section{Performance Evaluation}
\label{sec:evaluation}

\begin{figure*}[ht!]
    \centering
    \includegraphics[width=0.26\linewidth]{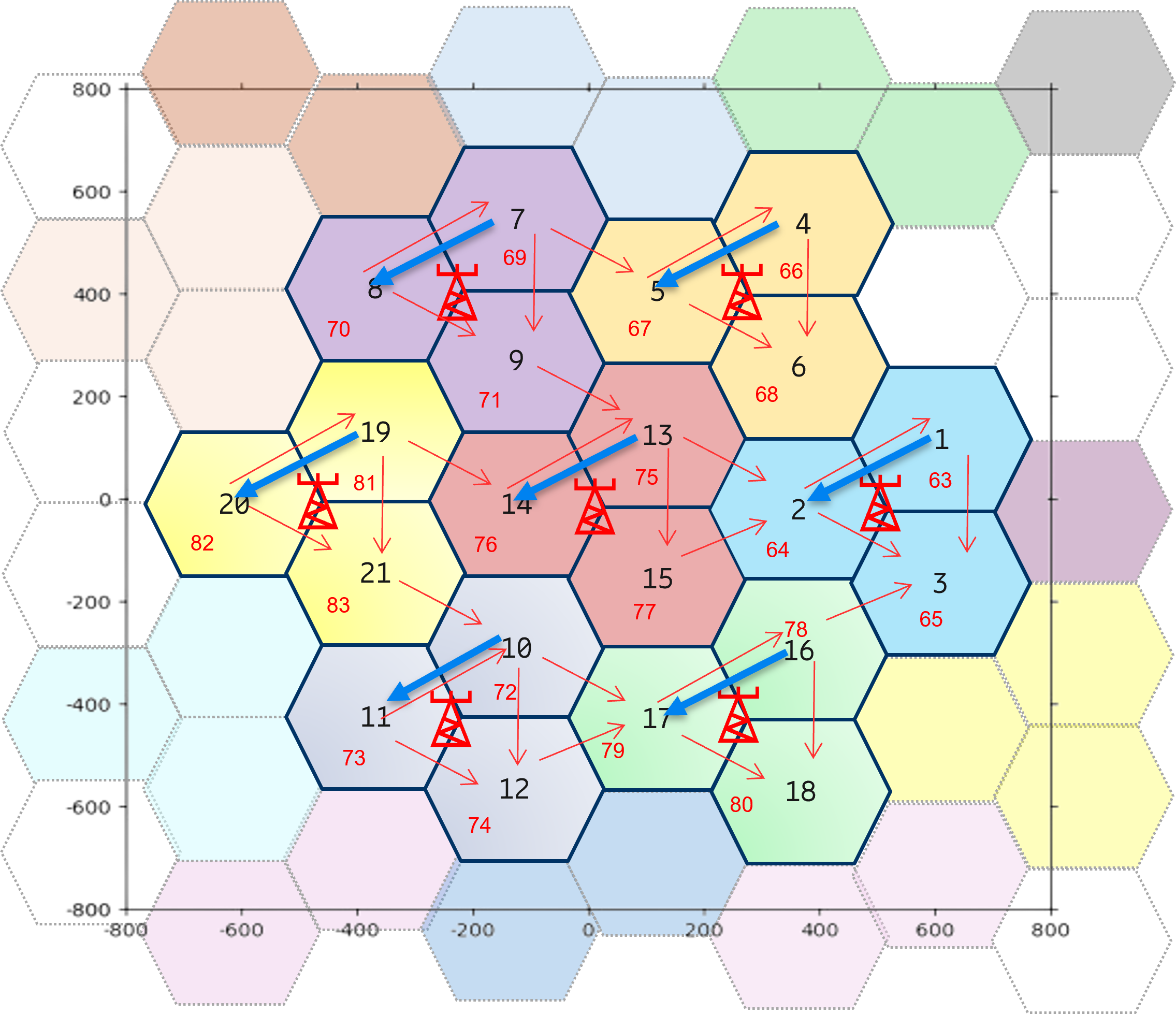}\includegraphics[width=0.31\linewidth]{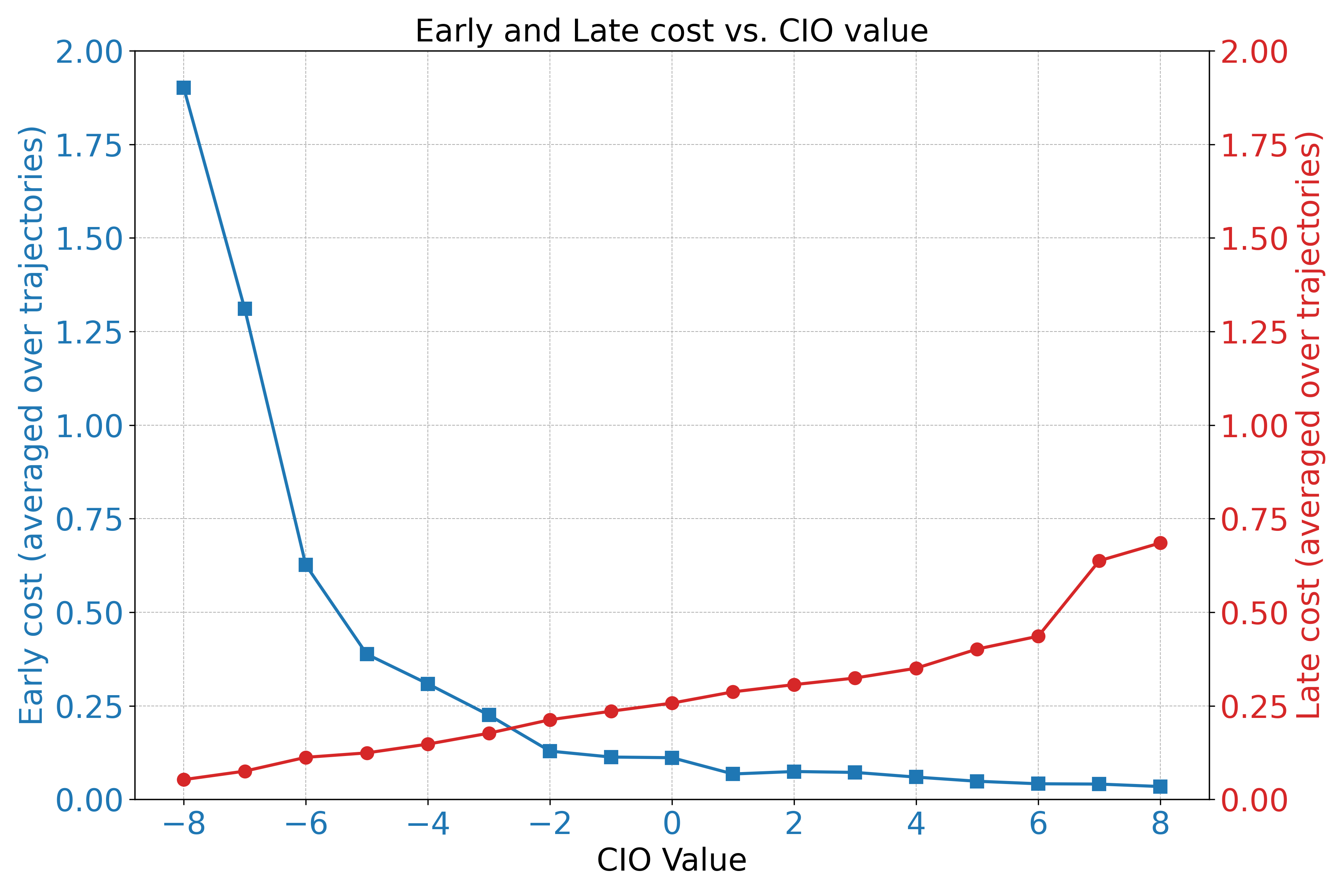}\includegraphics[width=0.34\linewidth]{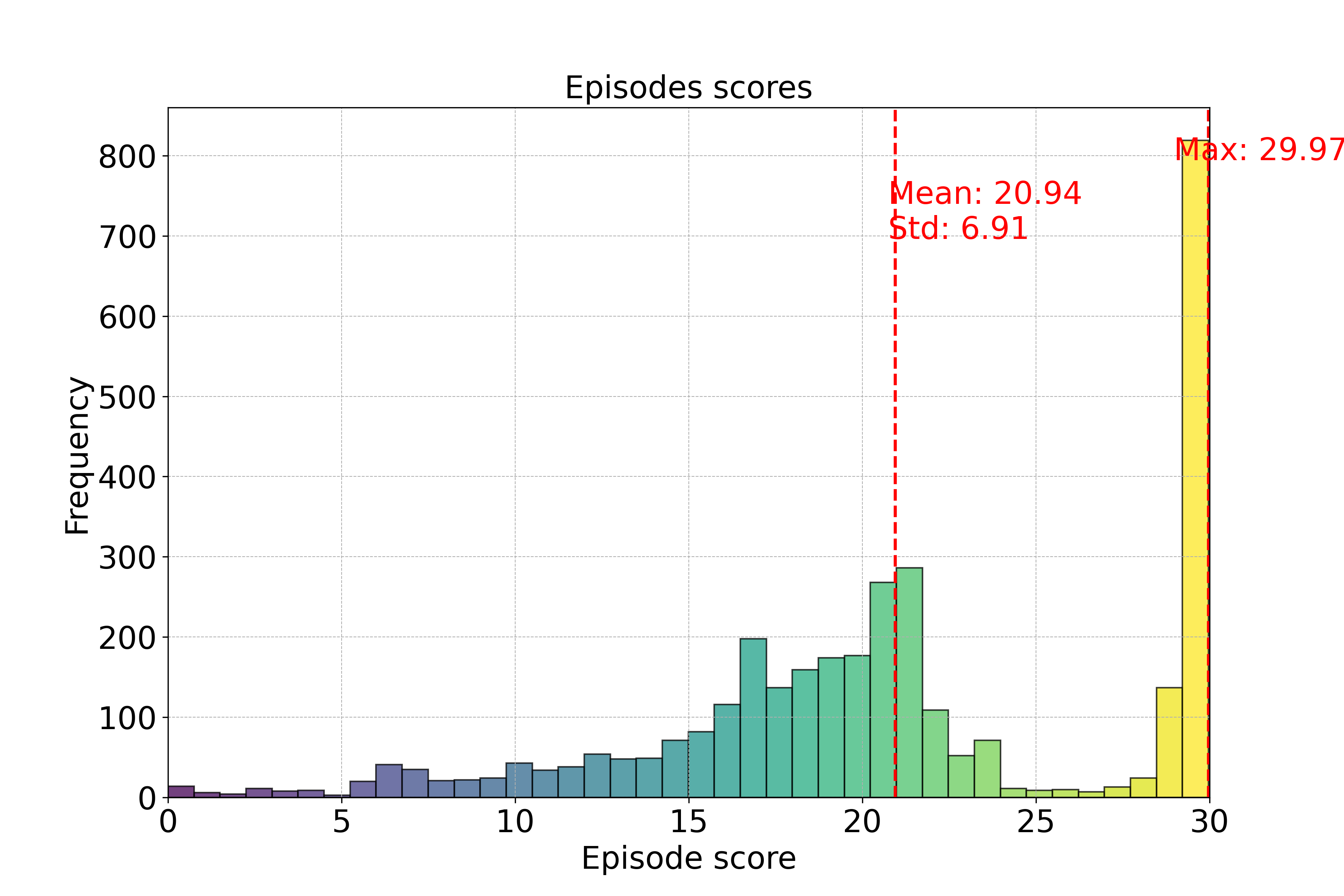}
    \caption{Simulated network (left), Plot of early and late event functions (middle), Distribution histogram of offline trajectory RtGs (right).}
    \label{fig:offline-data}
\end{figure*}


\begin{figure*}[t!]
    \centering
    \includegraphics[width=0.34\linewidth]{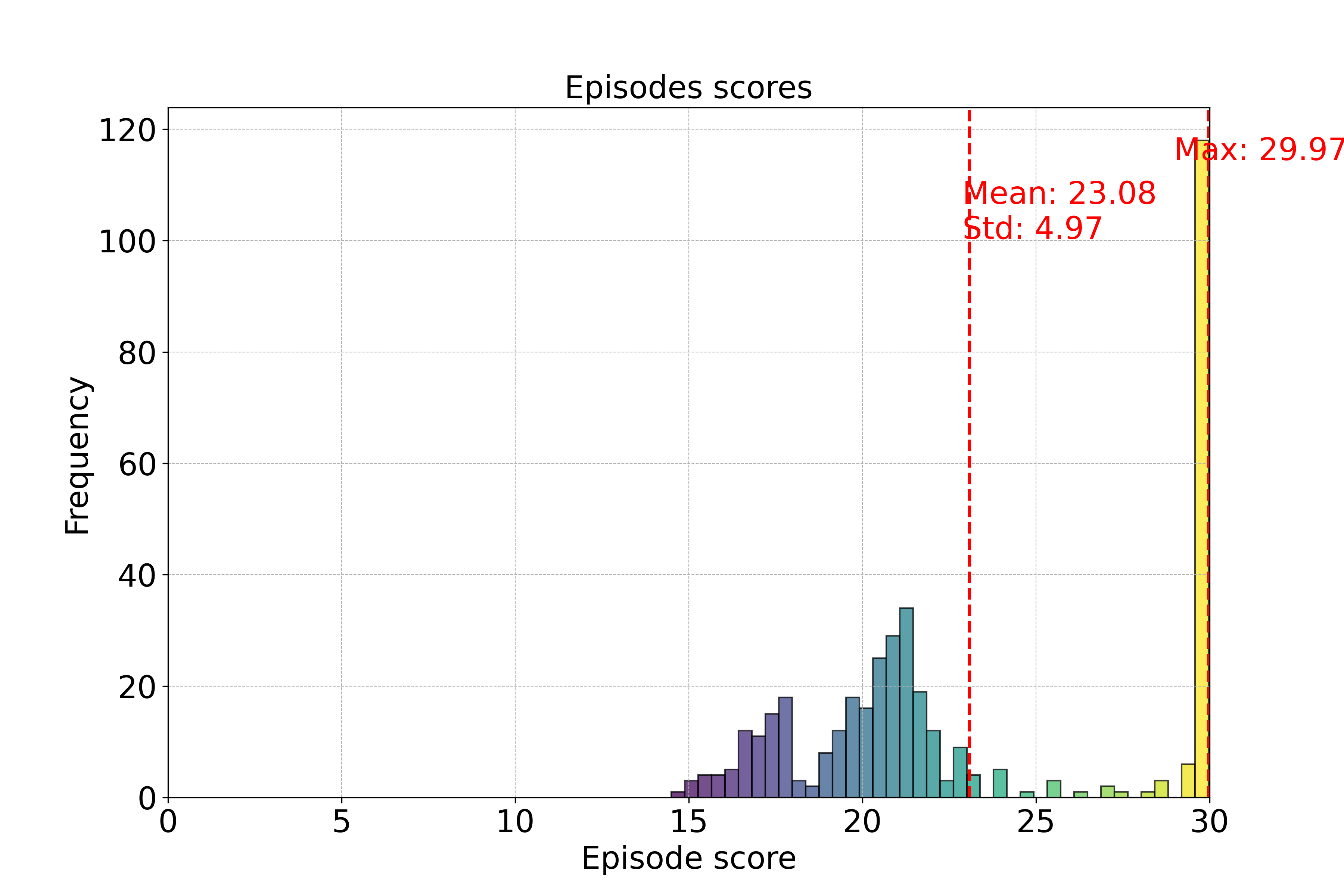}\includegraphics[width=0.34\linewidth]{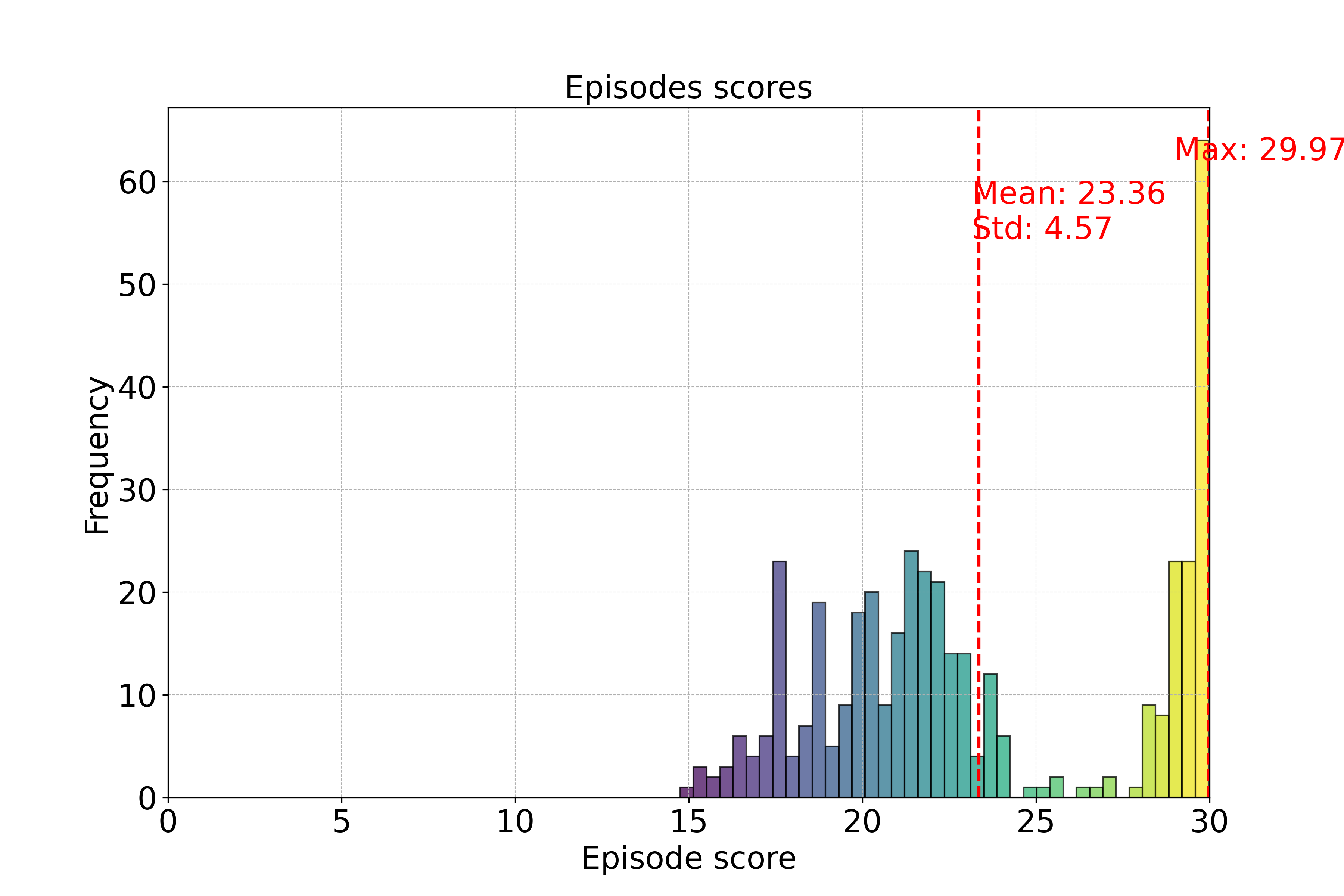}\includegraphics[width=0.30\linewidth]{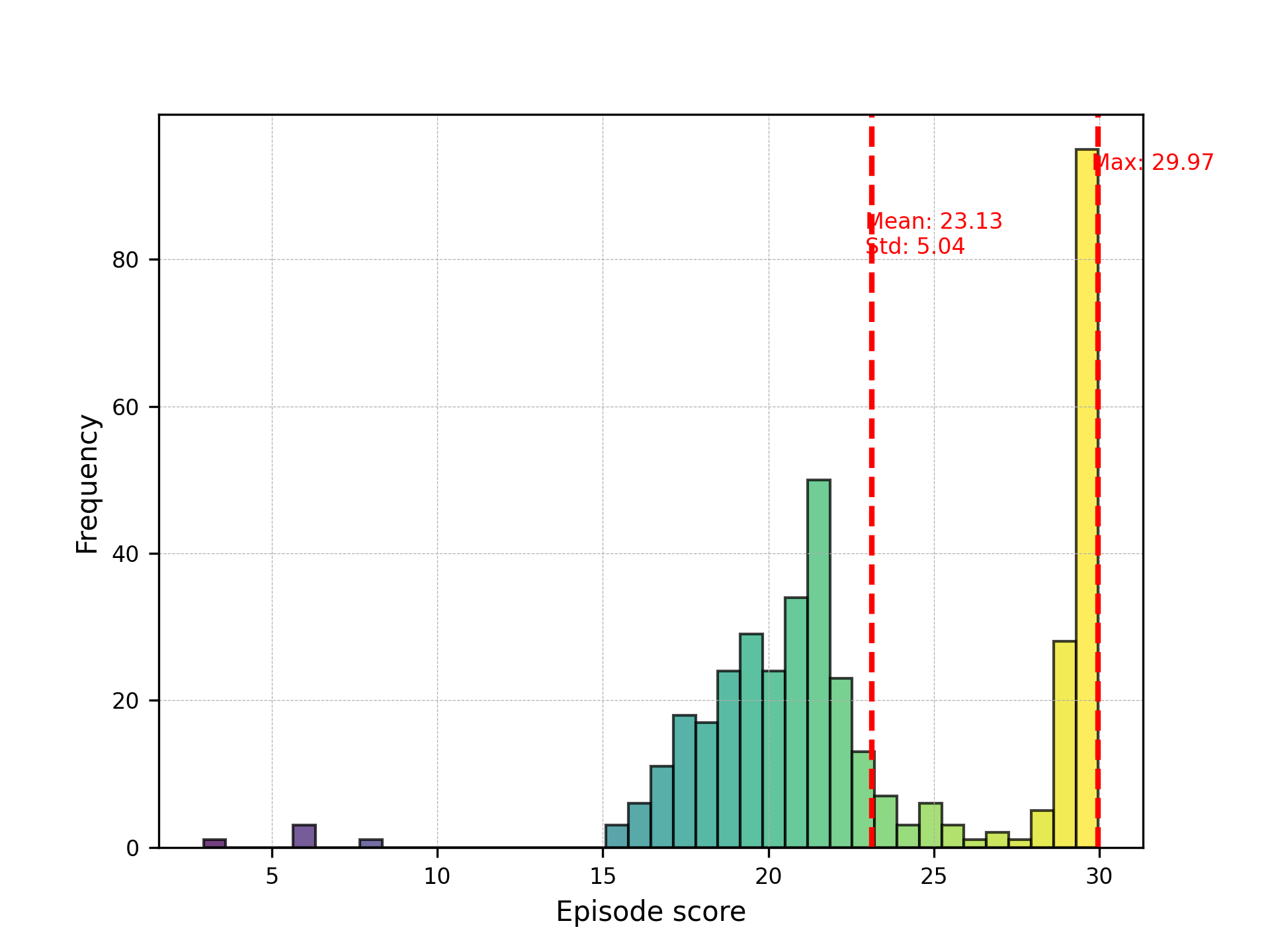}

    \caption{Test on train dataset: MRO rule-based (rb) mean: $23.08$ (left), DT with context $K=4$ mean: $23.36$ (middle), CQL policy, mean: $23.13$ (right).}
    \label{fig:test-on-train-policies}
\end{figure*}

\textbf{Simulation setting.} 
The simulations of this study have been done using an event-based proprietary company simulator for L2 and L3 layers that incorporates UE mobility. A network with $7$ Base Stations (BS), each with $3$ symmetric cells is simulated as shown in Fig.~\ref{fig:offline-data} (left). In this study, we consider the mid-band frequency at $3500$ MHz. Following inequality (\ref{rsrp_a3}), we fix the global A3 parameters as $Off_{A3} = 3 db$, $Hys_{A3}=1db$ and $TTT = 160ms$ based on similar empirical configuration in real systems. For MRO tuning, we fix all CIO parameters to $0$ except a subset of them, which will be tuned using the proposed approaches. As shown in  Fig.~\ref{fig:offline-data} (left), we choose for example to tune jointly the pairs $CIO_{1,2},\ CIO_{4,5},\  CIO_{7,8},\ CIO_{10,11},\ CIO_{13,14},\ CIO_{16,17}$ and $CIO_{19,20}$, which due to symmetry can be grouped together. More complex joint CIO tuning is left for future study. 

HO successes and failures result from the UE mobility trajectories as a user physically moves away from one cell and towards another, during the simulation window. A rule categorizes a HO-failure as early or late based on the combined information from the type of Radio Link Failure (RLF) known to the cell (e.g. Random Access Failure, Out-Of-Sync), and the stage during the handover process at which the failure occurred (e.g. before/after measurement report, before/after HO command received, etc).

\textit{Simulated traffic scenarios.} The scenarios include a list of UE traffic types: Video-on-Demand (VoD), gaming, Augmented Reality (AR), and web browsing. Half of the UEs are outdoor mobile, while the other half are indoor stationary. Gaming UEs live throughout the simulation period; the other types are Poisson generated over time and exit the system once their service is completed. For maximum traffic load $\ell=1$ approximately $600$ UEs are generated and served from $3$ cells during the simulation. The UE velocity and the traffic load 
are defined before each simulation. Different simulation scenarios use a fraction of the total load, e.g. $\ell = 0.6$. Randomness is controlled by fixing the seed, which determines the initial UE positions and their movement directions. The total simulation time for each scenario is $300$ seconds. Table~\ref{tab:simulations} lists the set of parameters used for train and validation.

For each of these training scenarios, we collect trajectories with policies RND, UP, DOWN, MRO for a random initial CIO value. The duration of all trajectories is fixed and equal to $T=17$; such choice corresponds to the maximum number of steps for our incremental agent to change the CIO from $-8$ (smallest CIO) to $+8$ (largest CIO). 
Considering the collected offline dataset, Fig.~\ref{fig:offline-data} (middle) plots the resulting empirical functions of early (\ref{early-sum}) and late issues (\ref{late-sum}) with respect to tuned CIO. We observe that the early issue function is monotone decreasing and the late is monotone increasing, with an intersection point around $-3dB$. This point is approximately the correct tuning point, and it is reasonable because having set the global offset to $Off_{A3}=+3dB$, their sum $Off_{A3}+CIO^* = +3-3=0\ dB$. 
This operation point can vary depending on cell load and average velocity, however, we will explore such variations in the MRO evaluations. 
In Fig.~\ref{fig:offline-data} (right) we plot the distribution of trajectory RtGs from the available offline trajectories that we use for training. We observe a very large distribution of values, ranging from $RtG=0$ (RND), to almost $RtG=30$ (expert policies). The bulk of RtGs is around the mean $20.94$.
\begin{table}[t!]
\caption{Parameters used to simulate the scenarios in RH.}
    \label{tab:simulations}
    \centering
    \resizebox{0.45\textwidth}{!}{ 
    \begin{tabular}{|c||c|c|c|}
        \hline
          & Load & velocity (km/h) & Seeds \\ 
        \hline
        train &0.2, 0.4, 0.6, 0.7       & 4, 50, 120   & 1,2       \\ 
        validation &  0.5, 0.6, 0.7     & 25, 85       & 3,4       \\  
        \hline
    \end{tabular}
    }
\end{table}
\subsection{Performance on the train set}
\begin{figure}[th!]
    \centering
    \includegraphics[width=0.7\linewidth]{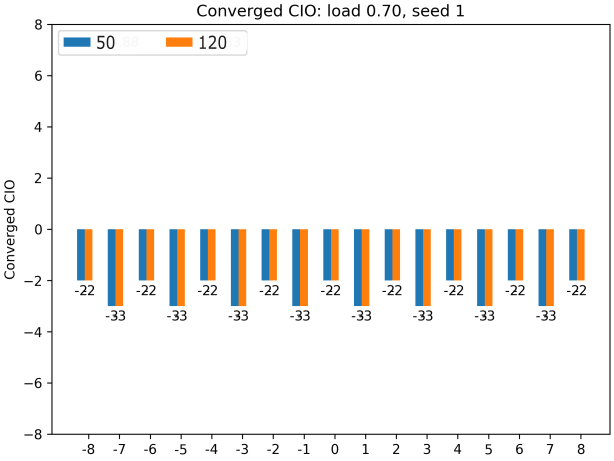}
    \caption{DT policy. Converged CIO value after $T$ steps (y-axis), for any initial starting CIO choice (x-axis), load $\ell=0.7$,  velocities $\left\{50, 120\right\}$ km/h.}
    \label{fig:DTconverge}
\end{figure}

As a first step we train both offline RL methods, i.e. DT and CQL, on the offline dataset and compare the learned models against the MRO rule-based (MRO-rb) on scenarios with same load and velocity as the train set (see Table~\ref{tab:simulations} first row). The results are detailed in Table~\ref{tab:test-on-train} and the RtG distributions from the three policies are illustrated in Fig.~\ref{fig:test-on-train-policies}. We observe that both
offline-RL policies manage to output trajectories with $RtG>15$ thus eliminating low reward RtGs from the dataset. MRO-rb is an expert-type of policy and has a similar behavior. We highlight however that the DT and CQL-fltr29 method can output on average a higher mean reward than MRO-rb. 
This implies that DT manages to "stich-together" parts of the available offline trajectories and produces trajectories with higher RtG than the MRO-rb, which was the highest performing policy in the dataset. Note, also that this is achieved for context length hyperparameter $K=4$, which is important to be appropriately tuned. We show in Table~\ref{tab:contextK} how DT performance varies with respect to this parameter. Another interesting result is illustrated in Fig.~\ref{fig:DTconverge}, where the x-axis shows the initial CIO choice at the beginning and the y-axis the converged CIO at the end of the $T=17$ length episode, following the learned DT policy, for load $\ell=0.7$ and two velocities $50$ and $120$ $km/h$. We observe that irrespective of the initial choice, the converged CIO value is $-3$ or $-2$, for both velocities. This indicates that the DT algorithm coherently tunes the CIO depending on the traffic conditions and not the initial CIO choice. The MRO-rb converged value is $-2$ for $50$ km/h
and $\left\{-3,-2\right\}$ for $120$ km/h, which explains the small improvement in mean performance due to the slightly modified tuning by the DT. 

The case of CQL is more complex. When learning directly CQL from the dataset, this does not perform well as shown in the third line of Table~\ref{tab:test-on-train}. The reason is that several trajectories in the dataset have high RtGs as a result of zero failures. But this lack of failures could be due to the short simulation time of $300$ sec, which confuses the CQL outcome. When we remove by hand these falsly high RtG trajectories, the resulting CQL-fltr29 can learn almost as good a policy as the DT method. Furthermore, as shown in the table for specific load and velocity scenarios $\ell=0.7, v=50$ and $\ell=0.4, v=50$ offline-RL (DT) can have as high as $+6\%$ relative performance improvement (rGain), due to slightly different tuning of the CIO value compared to MRO-rb.

On the validation set with unseen speeds $25\ km/h$ and $85$ $km/h$ we observe in Table~\ref{tab:test-on-val} that both CQL-fltr29 and DT obtain very good results for load $\ell=0.7$ outperforming MRO-rb by $+4\%$ and $+7\%$ respectively. However, for unseen load in training $\ell=0.5$ and unseen speeds we observe that DT performs $-1\%$
 lower than MRO-rb on average, whereas CQL-fltr29 still outperforms MRO-rb by $+4\%$. We can conclude that DT is prone to perform well as long as the environment is not very different than the one used in training, whereas CQL can generalize better to unseen scenarios. 
\begin{table}[t]
\caption{Test on train set: speeds $4,\ 50,\ 120$ km/h.}
\centering
\resizebox{0.48\textwidth}{!}{ 
\begin{tabular}{| c | c | c | c | c | c | c|}
\hline
Policy & load/v & mean & std & $CIO_{T}$ & rGain $\%$\\
\hline
RND & $all$ & $18.65$ & $7.50$ & - & $-19.2$ \\\hline
MRO-rb & $all$ & $23.08$ & $4.97$ & - & - \\\hline
CQL & $all$ & $20.69$ & $7.42$ & - & $-10.3$ \\\hline
CQL-fltr29 & $all$ & $23.13$ & $5.04$ & - & $\mathbf{+0.2}$ \\\hline
DT & $all$ & $\mathbf{23.36}$ & $\mathbf{4.54}$ & - & $\mathbf{+1.2}$ \\\hline\hline
MRO-rb & $0.7/50$ & $20.44$ & $0.70$ & $-2$ & -  \\\hline
DT & $0.7/50$  & $21.28$ & $0.77$ & $-3,-2$ & $\mathbf{+4.1}$\\\hline
MRO-rb & $0.4/50$ & $20.44$ & $0.70$ & $-3$ & -  \\\hline
DT & $0.4/50$  & $21.68$ & $0.99$ & $-4$ & $\mathbf{+6}$\\\hline
\end{tabular}
}
\label{tab:test-on-train}
\end{table}

\begin{table}[ht!]
\centering
\caption{Hyperparameter $K$ Tuning for DT.}
\resizebox{0.48\textwidth}{!}{ 
\begin{tabular}{| c | c | c | c | c | c | c|}
\hline
Policy & load/v & $K$ & mean & std & rGain $\%$\\
\hline
DT & $all$ & $3$ & $23.03$ & $4.61$ & $-0.2$ \\\hline
DT & $all$ & $4$ & $\mathbf{23.36}$ & $4.54$ & $\mathbf{+1.2}$ \\\hline
DT & $all$ & $5$ & $23.22$ & $4.51$ & $+0.6$ \\\hline
DT & $all$ & $7$ & $23.14$ & $4.64$ & $+0.2$ \\
\hline
\end{tabular}
}
\label{tab:contextK}
\end{table}

\begin{table}[t]
\centering
\caption{Test on validation set: speed $25$ km/h, $85$ km/h, unseen load $0.5$.}
\resizebox{0.48\textwidth}{!}{ 
\begin{tabular}{| c | c | c | c | c | c | c|}
\hline
Policy & load & mean & std & max & rGain $\%$\\
\hline
MRO-rb & $0.5$ & $38.1$ & $10.7$ & $60$ & - \\\hline
CQL-fltr29 & $0.5$ & $\mathbf{39.81}$ & $11.50$ & $66.37$ & $\mathbf{+4}$ \\\hline
DT & $0.5$  & $37.62$ & $10.34$ & $63.68$ & $\mathbf{-1}$\\\hline\hline
MRO-rb & $0.7$ & $44.27$ & $10.12$ & $58.1$ & -  \\
\hline
CQL-fltr29 & $0.7$ & $\mathbf{47.64}$ & $15.78$ & $89.13$ & $\mathbf{+7}$  \\
\hline
DT & $0.7$ & $\mathbf{46.05}$ & $11.69$ & $89.13$ & $\mathbf{+4}$ \\
\hline
\end{tabular}
}
\label{tab:test-on-val}
\end{table}

\begin{table}[ht!]
\caption{Test on train set for reward with CIO penalty.}
\centering
\resizebox{0.48\textwidth}{!}{ 
\begin{tabular}{| c | c | c | c | c | c | c|}
\hline
Policy & load/v & mean & std & $CIO_{T}$ & rGain $\%$\\
\hline
MRO-rb & $all$ & $21.36$ & $4.77$ & - & - \\\hline
DT & $all$ & $\mathbf{21.92}$ & $5.67$ & - & $\mathbf{+2.6}$ \\\hline\hline
MRO-rb & $0.7/50$ & $19.12$ & $0.83$ & $-2$ & -  \\\hline
DT & $0.7/50$  & $\mathbf{19.43}$ & $1.13$ & $\mathbf{-2,-1}$ & $\mathbf{+1.6}$\\\hline
MRO-rb & $0.4/50$ & $18.75$ & $1.13$ & $-3$ & -  \\\hline
DT & $0.4/50$  & $\mathbf{19.75}$ & $1.17$ & $\mathbf{-2,-1}$ & $\mathbf{+5.3}$ \\\hline
\end{tabular}
}
\label{tab:test-on-train-cio}
\end{table}
The last group of experiments considers the alternative reward function with CIO penalty, in (\ref{p-cio}). We can see in Table~\ref{tab:test-on-train-cio} that again offline-MRO (specifically DT) outperforms MRO-rb. In fact the performance can reach up to $+5.3\%$ higher than MRO-rb for load $\ell=0.4$ and velocity $50\ km/h$. The most important observation is that the offline RL converges to CIO values closer to $0$ compared to MRO-rb because of the soft-penatly term in the reward. 

\section{Conclusions}
\label{sec:conclusions}

This study demonstrates that offline RL trained on collected offline datasets can outperform traditional rule-based MRO methods in terms of reducing failures and other issues related to handovers. We have observed improvements up to $+7\%$ in previously unseen environments. 
Both DT and CQL methods showcase high performance, with DT performing slightly better for scenarios close to train data, and CQL on those further away. Using the same offline data our method can be trained with various types of rewards, 
thus offering higher flexibility compared to the rule-based MRO. This study has focused on tuning a unique source-target cell-pair for handovers. As a next step we plan to extend the RL study to account for the joint tuning and coordination of all cell-pairs in the network and differentiate tuning between UEs based on their service type.


\bibliographystyle{IEEEtran}
\bibliography{HO_bib}

\end{document}